\documentclass[conference]{IEEEtran}
\IEEEoverridecommandlockouts
\usepackage{cite}
\usepackage{amsmath,amssymb,amsfonts}
\usepackage{algorithmic}
\usepackage{graphicx}
\usepackage{textcomp}
\usepackage{xcolor}
\usepackage{graphicx}
\usepackage{etoolbox}
\usepackage{caption}
\usepackage{soul}
\usepackage{float}
\newcommand{\insertfig}
{ \setcounter{figure}{0}\includegraphics[width=0.9\textwidth]{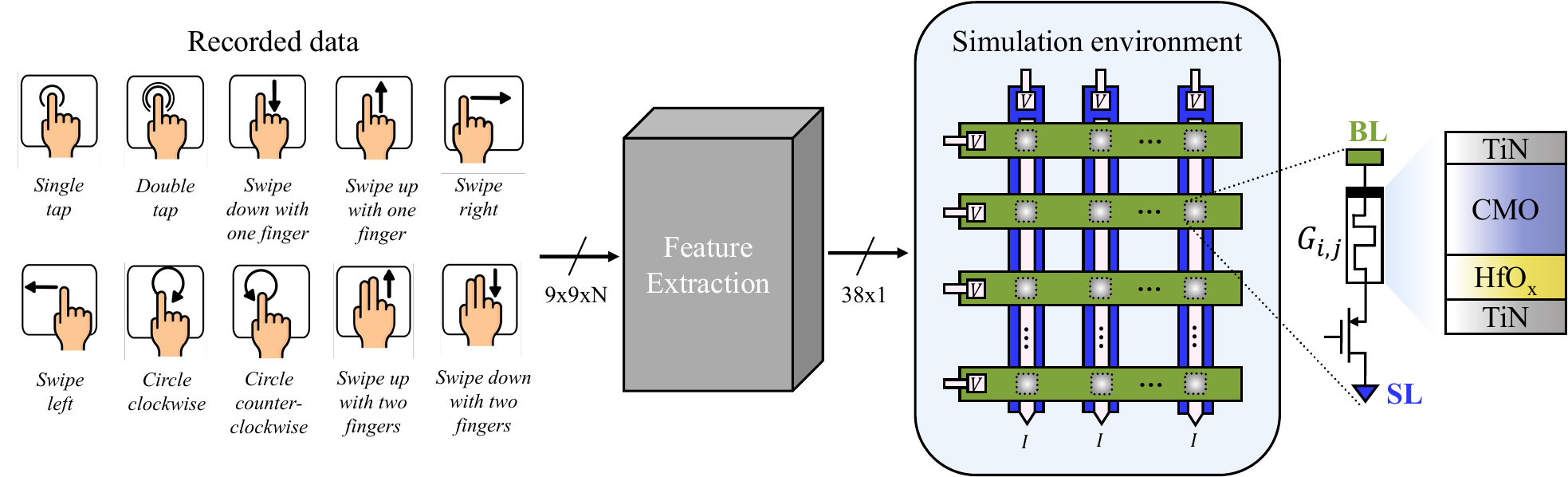}\captionof{figure}{\small Starting from the left, the schematic shows the ten hand gestures recorded on a textile-based tactile interface. Based on pressure, coverage area, and duration, a feature extraction method reduces the data space from $9\times 9\times N$ (where $N$ is the gesture duration) to 38$\times$1. These embeddings are then fed into a crossbar array on a simulation environment ("aihwkit"), calibrated using experimental data from a Conductive-Metal-Oxide (CMO)/HfO$_{\text{x}}$ analog ReRAM device stack, with TaO$_{\text{x}}$ acting as CMO layer.}\label{fig:intro}}
\def\BibTeX{{\rm B\kern-.05em{\sc i\kern-.025em b}\kern-.08em
    T\kern-.1667em\lower.7ex\hbox{E}\kern-.125emX}}
\makeatletter
\apptocmd{\@maketitle}{\centering\insertfig}{}{}
\makeatother
\begin{document}
\title{Edge Training and Inference with Analog ReRAM Technology for Hand Gesture Recognition\\
\thanks{This research is co-funded by the European Union and Swiss state secretariat SERI within the \textit{PHASTRAC} (grantID:101092096) project and by SNSF \textit{ALMOND} (grantID:198612).}
}

\author{
\IEEEauthorblockN{Victoria Clerico\textsuperscript{1},
Anirvan Dutta\textsuperscript{2}, Donato Francesco Falcone\textsuperscript{1},
Wooseok Choi\textsuperscript{1}, Matteo Galetta\textsuperscript{1},
Tommaso Stecconi\textsuperscript{1},\\András Horváth\textsuperscript{3},  Shokoofeh Varzandeh\textsuperscript{2}, Bert Jan Offrein\textsuperscript{1}, Mohsen Kaboli\textsuperscript{2}, and Valeria Bragaglia\textsuperscript{1,*}}\\

\IEEEauthorblockA{
    \textsuperscript{1}IBM Research Europe-Zürich, Switzerland \\
    \textsuperscript{2}BMW Group Research, RoboTac Lab, Germany \\
    \textsuperscript{3}Peter Pazmany Catholic University, Hungary \\
    *Email: vbr@zurich.ibm.com\\
  }
}
\maketitle
\captionsetup{font=small}

\begin{abstract}

Tactile hand gesture recognition is a crucial task for user control in the automotive sector, where Human-Machine Interactions (HMI) demand low latency and high energy efficiency. This study addresses the challenges of power-constrained edge training and inference by utilizing analog Resistive Random Access Memory (ReRAM) technology in conjunction with a real tactile hand gesture dataset. By optimizing the input space through a feature engineering strategy, we avoid relying on large-scale crossbar arrays, making the system more suitable for edge deployment. Through realistic hardware-aware simulations that account for device non-idealities derived from experimental data, we demonstrate the functionalities of our analog ReRAM-based analog in-memory computing for on-chip training, utilizing the state-of-the-art Tiki-Taka algorithm. Furthermore, we validate the classification accuracy of approximately 91.4\% for post-deployment inference of hand gestures. The results highlight the potential of analog ReRAM technology and crossbar architecture with fully parallelized matrix computations for real-time HMI systems at the Edge.
\end{abstract}


\section{Introduction}
Efficient and robust human-vehicle interfaces are crucial to enable intuitive interactions between user and machine, particularly in advanced driver-assistance systems \cite{murali2022intelligent}. The challenge lies in accurately interpreting subtle human communication cues, such as facial expressions, gestures, vocal tones, and body language, while ensuring prompt and precise system responses, critical for safe and enhanced user experience. However, achieving seamless integration requires optimizing computing latency and energy consumption, as real-time processing must maintain high performance.

This work focuses on tactile gesture recognition, where the decision relies on hand or finger motion on a touch interface \cite{braun2009using,touch_interface}. A promising solution inspired by the sensing mechanism of the human skin is textile-based tactile interfaces, which can be applied to large areas, complex shapes, and various resolutions \cite{bio_textile}. This technology requires efficient data processing, often resulting in high-overhead solutions. As haptic data increases in volume and complexity, traditional methods struggle to keep up \cite{Mehonic2022Growing}, leading to the use of Artificial Neural Networks (ANNs) to classify hand gesture patterns \cite{protractor,raksearch12,rnn_gesture_21}. Recent work evaluated traditional hand-gesture recognition approaches on such textile-based touch interface \cite{Fumelli2024}, highlighting the need for methods that can operate accurately in real-timez while considering energy efficiency. \\
\indent Analog in-Memory Computing (AiMC), a promising and energy-efficient approach for ANN accelerators, allows for executing highly parallel Multiply-Accumulate (MAC) operations—the most common computation in ANNs—within a single resistive crossbar array. This approach minimizes memory access and can hence be more power-efficient than traditional von-Neumann architectures. Computations are carried out directly within the hardware that stores the synaptic weights as conductance levels in resistive devices. Thus, MAC operations consist of applying voltage-encoded activation vectors and measuring the resulting currents. This approach has demonstrated up to 140$\times$ greater energy efficiency compared to digital accelerators when executing ANN inference tasks \cite{Jain2023Heterogeneous}. Among various emerging technologies, Resistive Random Access Memory (ReRAM) has gained significant attention over the past years for deep learning workloads. Its ability to store multiple non-volatile states, bidirectional accumulative updates, long retention, and fast switching speeds are key features to enable ANN training and inference acceleration in the analog domain. Several works have shown the potential of AiMC architecture based on ReRAM crossbar arrays for different applications \cite{Sebastian2020Memory, choi2021hardware, ReRAM1}, demonstrating the co-integration with 14nm CMOS technology \cite{Gong2022Deep}. \\
\indent ReRAM devices exhibit several non-ideal behaviors. Stochastic device-to-device variability, weight quantization—determined by the number of material states—and the non-linear asymmetrical responses of the devices pose challenges for on-chip inference and training. Optimization methods like Stochastic Gradient Descent (SGD) struggle with the inherent variability of resistive switching devices. Hence, specialized training algorithms like Tiki-Taka (TT) have been developed to account for memristor characteristics and improve convergence and performance \cite{Gong2022Deep, Rasch2024Fast}.
In this work, we perform hardware-level simulations demonstrating the potential of analog ANNs using ReRAM technology for efficient hand gesture recognition at the edge. The choice of ReRAM devices relies on their analog capabilities and improved properties compared to the state-of-the-art. Specifically ReRAM technology involving a bilayer stack of CMO/HfO$_\text{x}$ is taken as a reference, showing switching behavior in an analog fashion thanks to the presence of the CMO layer \cite{Stecconi2022Filamentary}. The simulation setup utilizes experimental data from Stecconi et al. \cite{Stecconi2024} to account for the device's non-idealities. To achieve this, we first preprocess the data to reduce the input space complexity, which relaxes the size requirements of the crossbar arrays for training and inference. We assess the performance of architectures with varying complexities, prioritizing shallow architecture suitable for edge applications and demonstrating the versatility and efficiency of our ReRAM technology for both training and inference.

\section{Data collection}
The touch interface used in this study for the hand gesture recognition task is the TexYZ textile-based capacitive sensor \cite{TeXYZ}, created through an embroidery process. This advanced tactile surface employs mutual capacitance detection to detect finger interactions on a 9$\times$9 taxel grid. It can be easily integrated into any surface for enhanced tactile functionality \cite{TeXYZ}. 
To accurately assess hand gesture recognition methods, we derived ten unique gestures \cite{Fumelli2024}: (1) \textit{Single tap}, (2) \textit{Double tap}, (3) \textit{Swipe down with one finger}, (4) \textit{Swipe up with one finger}, (5) \textit{Swipe right}, (6) \textit{Swipe left}, (7) \textit{Circle clockwise}, (8) \textit{Circle counter-clockwise}, (9) \textit{Swipe up with two fingers}, (10) \textit{Swipe down two fingers}, see Fig. \ref{fig:intro}.

Subsequent data collection sessions were conducted with 34 participants, each instructed to perform a set of gestures at three distinct speeds: regular, fast, and slow. Gestures were recorded continuously without time limits, allowing participants to perform at their natural pace for consistency and comfort. This approach produced variable-length gesture data, resulting in a total of 3060 time-series gestures. The collected data were then pre-processed using a running average filter, normalized for training, and augmented to increase the size of the dataset \cite{Fumelli2024}. The resulting dataset contains two forms of temporal (time-series) gesture data: raw pressure data and trajectory data (tracking the position of the finger over time). The pressure data consists of values recorded from each taxel in each time frame, yielding pressure data with dimensions of $9\times 9 \times N$ (where $N$ is the duration of the gesture). This time-series data was further processed and used as input vectors for the simulation of analog neural networks incorporating ReRAM experimental data.

\section{Methods}

\subsection{Data pre-processing}
We extracted features from tactile patterns to distinguish hand gestures within a constrained input space. Specifically, we derived key features from the pressure intensity, contact surface area, and gesture duration. Building on prior research \cite{tp_ref_1, tp_ref_2, Fumelli2024}, the features include mean and maximum pressure across all taxels over time. We computed pressure variability as the mean absolute difference between consecutive frames across all channels to capture dynamic changes. Gesture trajectory and peak count, defined by frames where the average pressure exceeds its neighbors, are also calculated to differentiate single from multi-finger gestures.

Spatial features include row-wise and column-wise mean pressure, representing the distribution of touch along the sensor array's length and width, respectively. Finally, we computed the contact area, considering the maximum and average values per frame. Thus, the input space was reduced from a $9\times9\times N$ format to a flat vector with a length of $38$ regardless of its duration, as indicated in Figure \ref{fig:intro}. For a more refined description of the feature definition refer to \cite{Fumelli2024}.s


\subsection{Hardware-level simulations}

The refined set of features is first used as input to the virtual crossbar arrays, representing the ANN layers within the simulation environment. The synaptic weights are then modeled according to experimental responses of the ReRAM devices from \cite{Stecconi2024}. These traces undergo a fitting procedure to reproduce the accumulative behavior of the devices. The cumulative response is crucial for accurately modeling the analog forward pass and weight updates, ensuring a faithful representation of device characteristics during inference and training.

\begin{figure}[h]
\centerline{\includegraphics[width=0.45\textwidth]{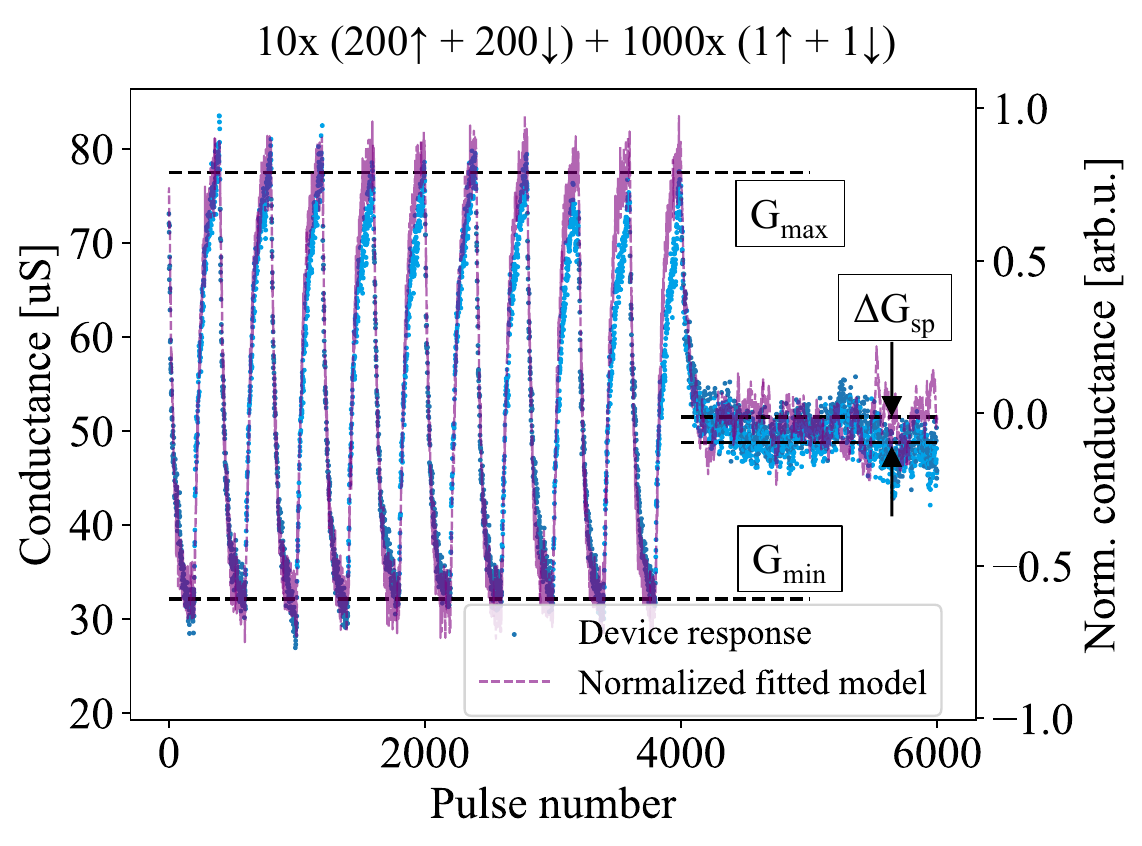}}
\caption{Experimental bidirectional accumulative response (blue) and its Soft-Bound model normalized fit (purple) obtained with a pulse scheme of 10 batches, each containing 200 identical positive (up) and negative (down) pulses followed by 1000 pulses with alternating polarity. Experimental data from \cite{Stecconi2024}.}\label{fig:pot_dep}
\end{figure}

Figure \ref{fig:pot_dep} shows a device bidirectional accumulative response to pulse trains with excellent graduality—essential for Tiki-Taka algorithm convergence \cite{Rasch2024Fast}—and minimal cycle-to-cycle variability throughout the 10 batches of up/down procedures. In this study, experimental data across 20 ReRAM devices were taken as a reference for the simulation, exhibiting a range of number of conductance states from 13 to 33, featuring a median of 22 \cite{Stecconi2024}.

To account for the conductance response of the ReRAM devices from \cite{Stecconi2024}, we made use of the “SoftBounds” (SB) model \cite{Frascaroli2018SoftBounds}, implemented in the open-source IBM platform ``aihwkit'' \cite{Rasch2021AIHWKIT}. This model replicates the conductance accumulative response of a ReRAM device (see Figure \ref{fig:pot_dep}) and allows for capturing cycle-to-cycle variations and the intrinsic skewness between increases and decreases in conductance from asymmetric switching processes.

We independently fitted each of the 20 devices by minimizing the average deviation between the SBs parametric model and the experimental data traces. 
Based on the tuned device models, a Gaussian multivariate distribution is created, where each device is defined by the number of states and the asymmetry between up and down responses, as described in \cite{Stecconi2024}. Both parameters were obtained from the fitted Soft-Bounds devices models, which considers an asymptotic behavior of the device saturation. This approach incorporates device-to-device variations into the simulation by sampling device characteristics from a Gaussian distribution that reflects the experimental data. Each device is sampled individually, leading to variations in the device responses, conductance ranges, and the number of states across synapses on the ANN layers. Hence, we provide a more hardware-aware and realistic scenario for simulating the analog ANN. 
\subsection{Analog Training and Inference setups}
We compared the performance of different architectures trained on the tactile gesture dataset features, with a train/test split of 75/25 . The analog ReRAM training result is compared to a Floating-Point (FP) 32-bit baseline neural network, trained using SGD, with the one of an analog training simulation that incorporates experimental data traces. 
The analog training simulation was carried out using Tiki-Taka version 2 (TTv2). This algorithm compensates for the inherent asymmetry in resistive switching devices by incorporating an additional matrix for gradient accumulation. To do so, the network's objective function is minimized along with an unintended cost introduced by asymmetrical behavior, ensuring more robust training on analog hardware \cite{TikiTaka}.

\begin{figure*}[t]
\centerline{\includegraphics[width=0.87\textwidth]{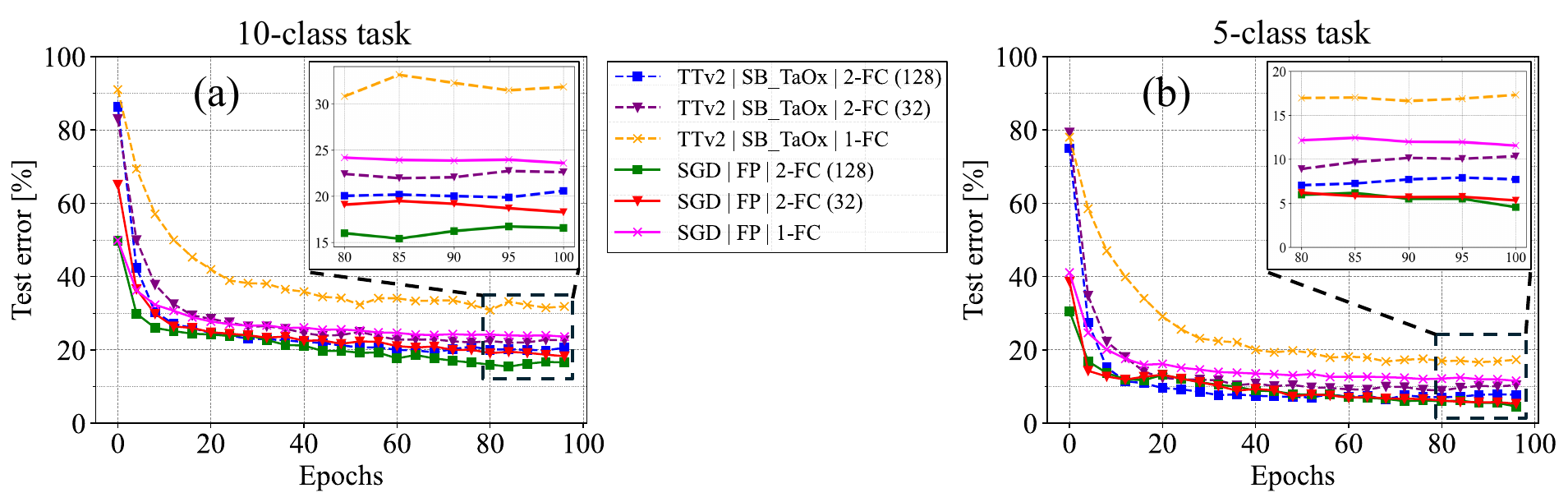}}
\caption{Training simulation of FC ANNs on (a) the full tactile hand gesture dataset and (b) the reduced 5-class set, comparing TTv2 analog (dashed) with 32-bit FP (solid). SB\_TaO$_\text{x}$ stands for the soft-bound fitted model of the TaO$_\text{x}$ devices. SGD training used $lr = 0.05$. TTv2 training used $fast\_lr$’ = 0.5, $transfer\_every = 5$, and $lr$ = 0.1 .}
\label{fig:training}
\end{figure*}

For inference, we simulated programming the ReRAM devices described in \cite{Stecconi2024} with conductances matching the pre-trained weights of a specified network. To define the FP baseline model, we performed hardware-aware tuning of the weights to inject noise and make the network more resilient to weight variations \cite{Rasch2023HWAware}. 
The devices are sampled from a Gaussian distribution based on the SBs-fitted device models, as in training. For inference, the fitted responses simulate conductance response upon pulses of alternating polarities. Then, to simulate the device programming, we followed the program-and-verify scheme (see Figure \ref{fig:inference}a). This algorithm applies voltage pulses to stimulate conductance changes until the target conductance is achieved within a tolerance. After simulating the program-and-verify scheme, the resulting conductances are used to instantiate synapses of an analog ANN using the "aihwkit" platform\cite{Gallo2023UseAIHWKIT}. Thereby, the inference assessment considers only programming non-idealities, excluding relaxation and read noise.

\section{Results}
\subsection{Training}
For our experimentation, we opted for Fully-Connected (FC) architectures since these are efficiently parallelized using analog crossbar arrays. We carried out the training of a 2-layer FC (2-FC) neural network to classify the 10 different hand gestures. We compared the performance per epoch for a 32-bit Floating Point (FP) reference trained using SGD (solid lines) and the accuracy when training on analog simulations using TTv2 (dashed lines), see Figure \ref{fig:training}a. The analog training simulation for 2-FC with 128 hidden neurons exhibits a 4.5\% accuracy drop compared to the FP baseline, achieving a classification accuracy of 81\%.

Additionally, we simplified the fully connected layers to just one (1-FC) to assess a more compact and shallow network. As a result, we observed a decrease in performance to 68\% in the analog training domain (see Fig.\ref{fig:training}b yellow and pink lines). However, by merging gestures with similar movements in opposite directions into one class, we reduced the task to five prominent gesture patterns - \textit{Tap}, \textit{Vertical swipe}, \textit{Horizontal swipe}, \textit{Circular}, and \textit{Two-finger swipe}. This selection was based on the most commonly used gestures in modern touch interfaces \cite{mitra2007gesture}. We then tested the simplified architectures on this simplified 5-class task, see Fig. \ref{fig:training}b. The simulation of analog training yielded an accuracy of 85.36\%, 3.2\% lower than the FP baseline. These results demonstrate the feasibility of performing in-situ training with Tiki-Taka using analog ReRAM for simplified tasks, shallow architectures, and reduced input spaces. This approach allows for fine-tuning the deployed array to adapt to different environments while utilizing the same crossbar architecture. 
\subsection{Inference}
\begin{figure}[h]
\centerline{\includegraphics[width=0.5\textwidth]{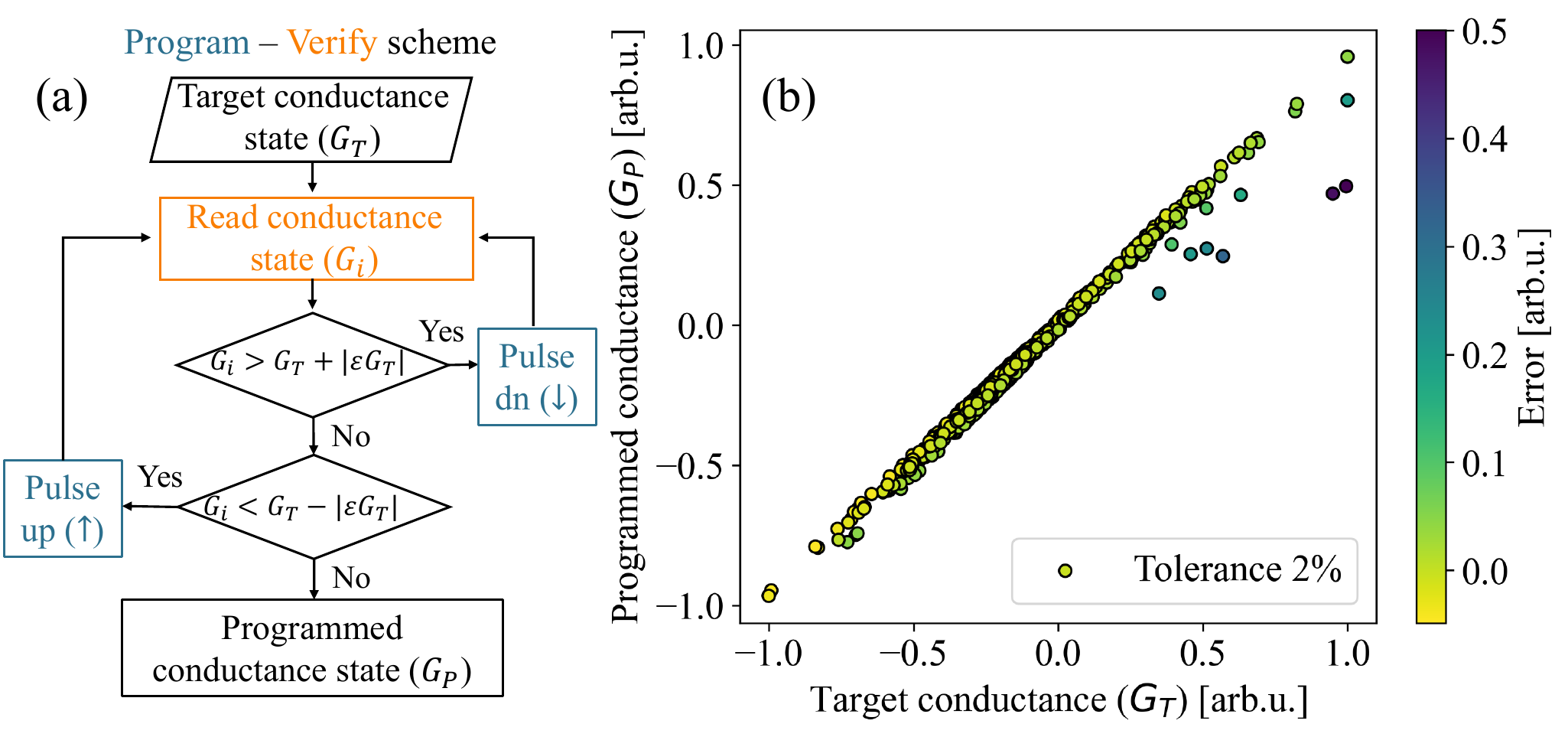}}
\caption{(a) Program-and-verify scheme for a target conductance $G_T$ and tolerance ($\varepsilon$). (b) Relationship between target and simulated programmed conductance on a normalized scale for $\varepsilon$ = 0.02.}
\label{fig:inference}
\end{figure}

For the inference evaluation, we selected the 2-FC network with 32 hidden neurons as the pre-trained model (see Fig. \ref{fig:training}b red line). We performed a simulation of the program-and-verify scheme depicted in Figure \ref{fig:inference}a for each device comprising the analog layers' synapses of the baseline network. First, we applied min-max normalization to the pre-trained weights to adjust the values to the same normalized conductance range. We performed the program-and-verify simulation considering a tolerance of 2\% of the target conductance and a maximum of 200 iterations. 

In the simulation, we programmed each device to a target conductance given by the pre-trained weight of its corresponding synapse. 
Our simulations showed high programming precision for the specified conductance ranges, see Figure \ref{fig:inference}b. After setting the device conductances to the pre-trained weights, we tested the inference accuracy on a test set on the tactile gesture data. The FP baseline showed a 94.28\% classification accuracy (red solid line from Figure \ref{fig:training}b) whereas the analog network showed a post-programming inference accuracy of 91.14\% on the testing set. These results show that precise programming boosts inference performance, underscoring the technology's versatility for both training and inference.
\section{Conclusion}
We presented Edge analog AI training and inference applications for tactile hand gesture recognition based on emerging analog ReRAM technology. Through feature extraction, we optimized the ReRAM-based neural network input space to deploy on small crossbar array configurations. We conducted on-chip learning simulations using realistic data from multiple devices, incorporating device-to-device variability.  Leveraging the state-of-the-art Tiki-Taka algorithm, the simulations achieved an accuracy of 85.36\%, a 3.2\% lower than the FP baseline. 
Moreover, we simulated an offline pre-trained Edge inference scenario with a program-and-verify scheme, demonstrating an accuracy of 91.14\%, a 3.16\% lower than the pre-trained FP baseline. These results highlight the effectiveness of analog ReRAM technology for efficient Edge AI training and inference, showcasing its impressive capabilities. This study paves the way for implementing on-chip Edge inference with analog ReRAM scalable crossbar arrays.
\section*{Acknowledgment}
The authors acknowledge the Binnig and Rohrer Nanotechnology Center (BRNC) at IBM Research Europe - Zurich. The authors would like to thank Malte J. Rasch for the technical discussions and Chiara Fumelli for the groundwork that laid the foundation for this study.
\bibliography{biblio}
\bibliographystyle{IEEEtran}

\end{document}